\pdfoutput=1
\documentclass[runningheads]{llncs}
\usepackage[T1]{fontenc}
\usepackage{graphicx}
\usepackage{caption}
\usepackage{xcolor}
\usepackage{hyperref}
\hypersetup{hidelinks,
    colorlinks=true,
    linkcolor=red,
    citecolor=black,
    urlcolor=blue,
	pdfstartview=Fit,
	breaklinks=true}
\usepackage{multirow}
\usepackage{url}
\begin{document}
\title{3D-EffiViTCaps: 3D Efficient Vision Transformer with Capsule for Medical Image Segmentation}
\titlerunning{3D-EffiViTCaps}
\author{Dongwei Gan \and Ming Chang \and Juan Chen}
\authorrunning{Dongwei Gan et al.}
%
\institute{School of Computer Science and Engineering, \\University of Electronic Science and Technology of China, Chengdu, China}
\maketitle              
\begin{abstract}
Medical image segmentation (MIS) aims to finely segment various organs. It requires grasping global information from both parts and the entire image for better segmenting, and clinically there are often certain requirements for segmentation efficiency. Convolutional neural networks (CNNs) have made considerable achievements in MIS. However, they are difficult to fully collect global context information and their pooling layer may cause information loss. Capsule networks, which combine the benefits of CNNs while taking into account additional information such as relative location that CNNs do not, have lately demonstrated some advantages in MIS. Vision Transformer (ViT) employs transformers in visual tasks. Transformer based on attention mechanism has excellent global inductive modeling capabilities and is expected to capture long-range information. Moreover, there have been resent studies on making ViT more lightweight to minimize model complexity and increase efficiency. In this paper, we propose a U-shaped 3D encoder-decoder network named 3D-EffiViTCaps, which combines 3D capsule blocks with 3D EfficientViT blocks for MIS. Our encoder uses capsule blocks and EfficientViT blocks to jointly capture local and global semantic information more effectively and efficiently with less information loss, while the decoder employs CNN blocks and EfficientViT blocks to catch finer details for segmentation. We conduct experiments on various datasets, including iSeg-2017, Hippocampus and Cardiac to verify the performance and efficiency of 3D-EffiViTCaps, which performs better than previous 3D CNN-based, 3D Capsule-based and 3D Transformer-based models. We further implement a series of ablation experiments on the main blocks. Our code is available at: \href{https://github.com/HidNeuron/3D-EffiViTCaps}{https://github.com/HidNeuron/3D-EffiViTCaps}.

\end{abstract}
\section{Introduction}
With the rapid advancement of medical technology, medical imaging has become an important reference information for doctors to understand and analyze conditions, and plays a significant part in diagnosing diseases and evaluating treatment. Medical image segmentation (MIS) is aimed at identify the pixels of diseased organs in medical images with the objective of extracting the information features of these diseased areas. Apart from traditional segmentation methods, with the rapid growth of computer hardware performance, deep learning approaches demonstrate significant potential in image processing tasks. U-Net \cite{2015U}, proposed in 2015, has received widespread attention due to its scalable structure and exceptional performance in segmenting tiny objects. With the rising demand for MIS performance in recent years, numerous academics have been constantly upgrading and expanding the U-Net structure, such as extending U-Net to 3D images \cite{2016V}, \cite{iek20163D}, as well as the feature fusion methods \cite{2018UNet}, \cite{2019MultiResUNet}, etc. Nevertheless, due to the shortcomings of CNNs such as the difficulty of capturing global context information, sensitivity to rotation and affine transformation \cite{20223D}, the loss of translation invariance and a certain amount of information caused by the pooling layer, etc., the performance of networks relying on CNNs is subject to certain limitations.

To address the possible limitations of CNNs, Sabour et al. \cite{2017Dynamic} proposed capsule network. It was primarily suggested to address the issue of convolutional neural networks overlooking the inter-feature relationships during information extraction. Capsule is a composite vector with multiple neuron scalars. It can identify a visual entity and output the direction, size, relative position between objects, etc. of the entity. The core algorithm of the capsule network is the dynamic routing algorithm, which extracts feature information while not discarding other relevant information. Initially, capsule was predominantly utilized for image recognition and demonstrated promising outcomes. However, in terms of semantic segmentation, at the beginning, the performance of Capsule-based models was not very impressive because of the strong performance of CNN-based models. SegCaps \cite{2018Capsules}, \cite{2021Capsules}, a 2D encoder-decoder network, are the first capsule-based methods utilized in MIS. Although the effect is not significant, it has made a certain contribution to the use of capsule networks in MIS. However, as a 2D method, SegCaps lack the ability to process 3D data. Poor performance often results from the loss of some temporal information when utilized to segment volumetric images. In order to overcome the shortcomings of the 2D capsule network, Nguyen et al. \cite{20223D} proposed 3D-UCaps, whose encoder is based on the 3D capsule blocks and the decoder is composed of 3D convolutional blocks. Its performance turns out to be more robust than the 2D capsule network. Subsequently, Tran et al. \cite{2022SS} used self-supervised learning for 3D-UCaps and developed SS-3DCapsNet to solve the problem of the former being highly dependent on sampling order and weight initialization. For retaining the ability of CNNs to make good use of local information to model short-range attention and reduce the huge inference time caused by excessive capsule layers, Tran et al. \cite{20223DConvCaps} redesigned the encoder pathway and used convolutional layers for capturing low-level features while capsule blocks for handling high-level features. Compared with previous 3D models \cite{20223D}, \cite{2022SS}, 3DConvCaps \cite{20223DConvCaps} performs better and has shorter inference time.

Nevertheless, the local receptive fields of CNNs cannot capture global information well. Since transformer \cite{2017Attention}, which is based on the attention mechanism, was proposed, it has produced excellent results in the field of natural language processing (NLP). It has subsequently been applied to various modalities and achieved breathtaking performance. Vision Transformer (ViT) \cite{2020An} separates the input image into patches and projects each patch into a fixed-length vector before sending it to the transformer. The subsequent operations of the encoder are identical to those of the original transformer. Swin Transformer \cite{2021Swin1} was proposed based on ViT and improved upon it, using Patch Merging blocks to obtain multi-scale information and a novel way to compute the attention (i.e. window attention). Cao et al. \cite{2021Swin2} combined Swin Transformer with U-Net and proposed the Swin-Unet for MIS, designing the Patch Expanding operation in the decoder. Peiris et al. \cite{peiris2022robust} then proposed a 3D model called VT-UNet and applied it to 3D organ tumor segmentation. However, many operations of ViT, such as frequent reshape, element-wise addition, and normalization, are memory inefficient and require time-consuming access across different storage units. Recently, there has been a lot of work aimed at improving the efficiency of traditional ViT \cite{huang2022lightvit}, \cite{cai2022efficientvit}, \cite{liu2023efficientvit}. Liu et al. \cite{liu2023efficientvit} proposed EfficientViT, introducing cascaded group attention to optimize the inference efficiency of ViT. EfficientViT strikes a decent balance of efficiency and accuracy and surpasses existing efficient models.

Motivated by the success of 3D capsule network \cite{20223D}, \cite{2022SS}, \cite{20223DConvCaps} and 3D U-shaped transformer \cite{2021Swin2}, \cite{peiris2022robust} on MIS, as well as the improvement of model performance by EfficientViT \cite{liu2023efficientvit}, we propose 3D-EffiViTCaps to capture local and global information better and further improve model performance while keeping a balanced model efficiency. First, the input images are passed to the feature extraction block to initially extract features and then passed to the encoder. The 3D Patch Merging and 3D EfficientViT blocks capture low-level features, while the 3D capsule blocks handle high-level ones. The retrieved features are subsequently given to the decoder, which uses 3D convolutional and 3D EfficientViT blocks to restore the information to the input resolution to achieve pixel-level prediction. Our model balances efficiency and performance well, and it has improved segmentation accuracy. Our ablation study demonstrates the superiority of using the 3D EfficientViT blocks and the rationality of the settings of other blocks. In summary, our contribution can be summarized as: (1) While using 3D capsule to model the part-whole relations, 3D EfficientViT is utilized to better extract global semantic information and reduce information loss, which further increases the segmentation performance of the MIS model. (2) Maintain a plausible model size and calculation amount, achieving a decent balance between model performance and efficiency.

\section{Related Work}

\noindent\textbf{CNN-based MIS}: In the development process of MIS methods, there have been different traditional segmentation algorithms, mainly based on contour and machine learning. As deep neural networks continue to advance, CNNs have risen rapidly, gradually sweeping the entire computer vision world. U-Net \cite{2015U} was proposed to achieve MIS, and subsequently various U-Net-based segmentation models emerged endlessly, such as U-Net++ \cite{2018UNet}, U-Net 3+ \cite{2020UNet}, Non-local U-Net \cite{2020Non}. U-Net has also been introduced into the 3D MIS field, and many classic 3D models came out, such as 3D U-Net \cite{iek20163D} and V-Net \cite{2016V}, which have shown relatively robust performance. Despite the fact that CNN-based methods have been quite successful in the field of MIS because of their strong representation learning capabilities, they are subject to certain limitations such as capturing global information and long-range dependencies.

\noindent\textbf{Capsule-based MIS}: Similar to the CNN-based MIS methods, the Capsule-based MIS methods \cite{2018Capsules}, \cite{2021Capsules}, \cite{20223D}, \cite{2022SS}, \cite{20223DConvCaps} also build on an encoder-decoder structure. But different from CNNs, the capsule network \cite{2017Dynamic} encodes spatial information and also calculates the presence probability of the entity. Each layer of capsules will calculate its own activity vector based on the activity vector received from the previous layer capsules, and send it to the next layer capsules to learn a set of entities and represent them in a high-dimensional form. Mainly based on the dynamic routing algorithm, the simple feature in the previous layer make an agreement in their votes to choose whether or not to turn on the rather complicated feature in the next layer. Through the message passing method of capsule, the model can better learn the part-whole relationship.

SegCaps \cite{2018Capsules}, \cite{2021Capsules} are the initial Capsule-based models proposed for 2D MIS. Although there is a certain gap compared with powerful CNN-based models, they are a breakthrough for capsule networks in MIS. However, SegCaps were originally designed for 2D images. When they are used to process volumetric images, they do not perform very well due to the loss of temporal information. To apply capsule network to 3D MIS, Nguyen et al. \cite{20223D} proposed 3D-UCaps, with 3D capsule blocks in decoder and 3D convolutional blocks in decoder. Then, Tran et al. \cite{2022SS} used self-supervised learning based on 3D-UCaps and came up with SS-3DCapsNet to address the issue of the former being heavily reliant on weight initialization. Nevertheless, even if the Capsule-based blocks have fewer parameters, the model efficiency is still relatively low due to the dynamic routing mechanism, and it is not good at capturing short-range information. Tran et al. \cite{20223DConvCaps} proposed 3DConvCaps, which uses convolutional layers to capture low-level features and capsule blocks to encode part-whole relationship. However, even when a convolution with a stride of 2 is used in place of the pooling layer, there is still considerable information loss and CNNs do not perform well in capturing global attention information.  

\noindent\textbf{Vision Transformer in MIS}: Once the attention mechanism-based transformer \cite{2017Attention} was proposed, it has quickly attained state-of-the-art outcomes in numerous NLP tasks. In view of the huge success of transformer, many subsequent researchers designed a variety of attention mechanisms and applied transformer to MIS, most of which achieved very significant results, e.g. \cite{2021TransUNet}, \cite{jose2021medical}, \cite{2021TransFuse}. Swin-Unet \cite{2021Swin2} proposed by Cao et al. is the first U-Net-shaped MIS network based on pure transformer. Swin Transformer \cite{2021Swin1} is used to construct the encoder, bottleneck and decoder, and its performance is better than TransUnet \cite{2021TransUNet}, etc. Swin Transformer is mainly based on Vision Transformer (ViT) \cite{2020An}. The former also divides the input image into several patches like ViT, but is improved using the shifted window (SW) mechanism. Apart from the SW, it continuously increases the size of the patches through the Patch Merging blocks, so that multi-scale information can be obtained. Cao et al. \cite{2021Swin2} combined U-Net \cite{2015U} and Swin Transformer, and inspired by the Patch Merging operation, designed the Patch Expanding operation on the decoder pathway, then proposing Swin-Unet. Peiris et al. \cite{peiris2022robust} extended the Swin Transformer model based on U-Net architecture to 3D (named VT-UNet) and used it for 3D organ tumor segmentation. However, the speed of traditional transformer models is often limited by memory-inefficient operations and computational redundancy of attention maps, especially multi-head self-attention (MHSA). The recent EfficientViT \cite{liu2023efficientvit} adopts a "sandwich layout" strategy and proposes a cascaded group attention block, reduces computing expenses while also increasing attention diversity. Inspired by that, we modify the EfficientViT block into 3D and apply it to the MIS model, which improves the segmentation accuracy while achieving a decent balance between model performance and efficiency.

\section{Our Proposed 3D-EffiViTCaps}

\noindent\textbf{\large Main Network Blocks} 

\textbf{Capsule Block}: As mentioned above, capsule network \cite{2017Dynamic} mainly extract useful information in the area through dynamic routing algorithm, and almost does not discard any information. Using appropriate capsule blocks during downsampling can better capture part-whole relationship while reducing information loss. The capsule block integrates vectors containing low-level feature information into high-level feature vectors through weighted sum. Let $\{{c_{1}^{l}},{c_{2}^{l}},...,{c_{n}^{l}}\}$ represent the capsules of the layer $l$, and $\{{c_{1}^{l+1}},{c_{2}^{l+1}},...,{c_{m}^{l+1}}\}$ represent the capsules of the layer $l+1$. The overall dynamic routing process is formulated as:
\begin{equation}
c_{j}^{l+1}=squash(\sum\limits_{i}r_{ij}v_{j|i}), \quad v_{j|i}=W_{ij}c_{i}^{l}\label{eq}
\end{equation}
where $r_{ij}$ is the coupling coefficient between the $i$-th capsule in layer $l$ and the $j$-th capsule in layer $l+1$, which is obtained through dynamic routing algorithm and satisfies $\sum_{j}r_{ij}=1$. $W_{ij}$ is a weight matrix learned through backpropagation that maps low-level feature vectors to vectors in high-level feature space.

\begin{figure}[h]
  \setlength{\belowcaptionskip}{-2cm}
  \centering
  \includegraphics[width=\textwidth]{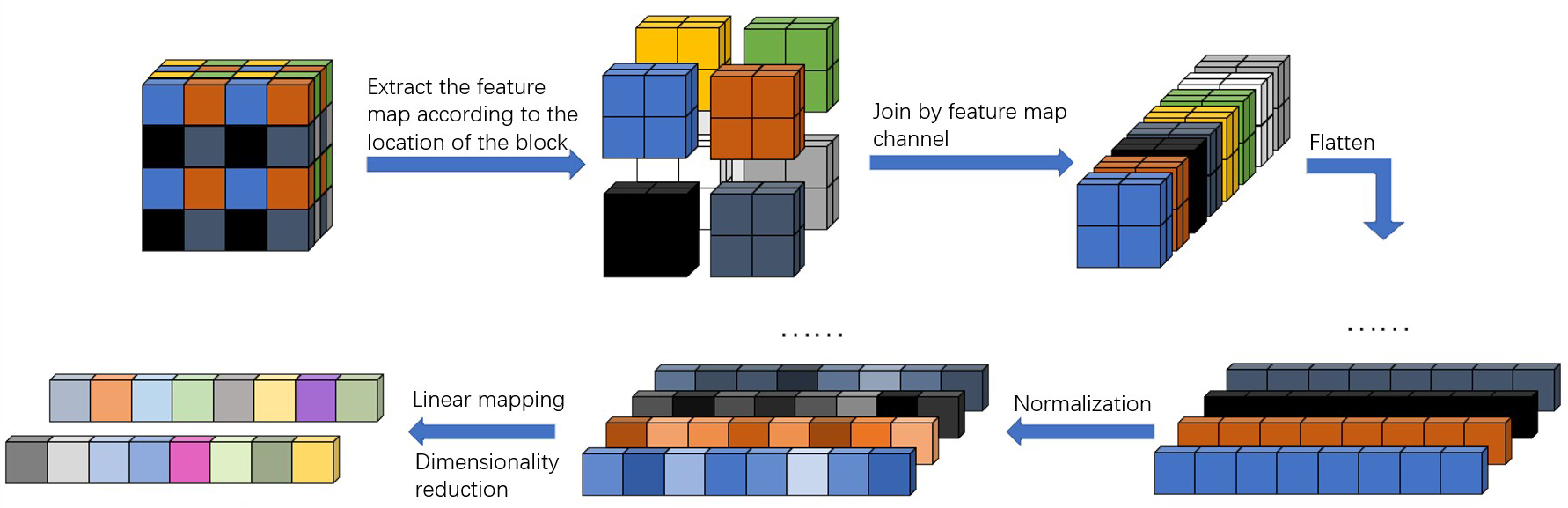}
  \caption{Principle Display of 3D Patch Merging}
  \label{Patch Merging}
\end{figure}

\textbf{3D Patch Merging Block}: We use the 3D Patch Merging blocks for downsampling to reduce the resolution and alter the channel dimension to create a hierarchical structure. It is similar to the pooling layer of CNNs or the convolutional layer with stride 2, but it loses almost no information. Furthermore, compared to the convolutional layer with stride 2, the 3D Patch Merging blocks save a certain amount of calculations and reduce the FLOPs count of the whole model while achieving a slightly better performance.

As shown in Fig.\ref{Patch Merging}, refer to the 3D Patch Merging block of VT-UNet \cite{peiris2022robust}, but the difference is that we select elements according to the position interval 2 in the height (H), width (W) and depth (D) directions (instead of only height and width) to form new patchs, then concatenate all the patches as a whole tensor, and finally expand it. The channel dimension will now increase to 8 times its initial size (because H, W, and D are each reduced to half of the original size). Afterwards, a fully connected layer is employed for projection to modify the channel dimension. 

\begin{figure}[h]
  \setlength{\belowcaptionskip}{-0.5cm}
  \centering
  \includegraphics[width=\textwidth]{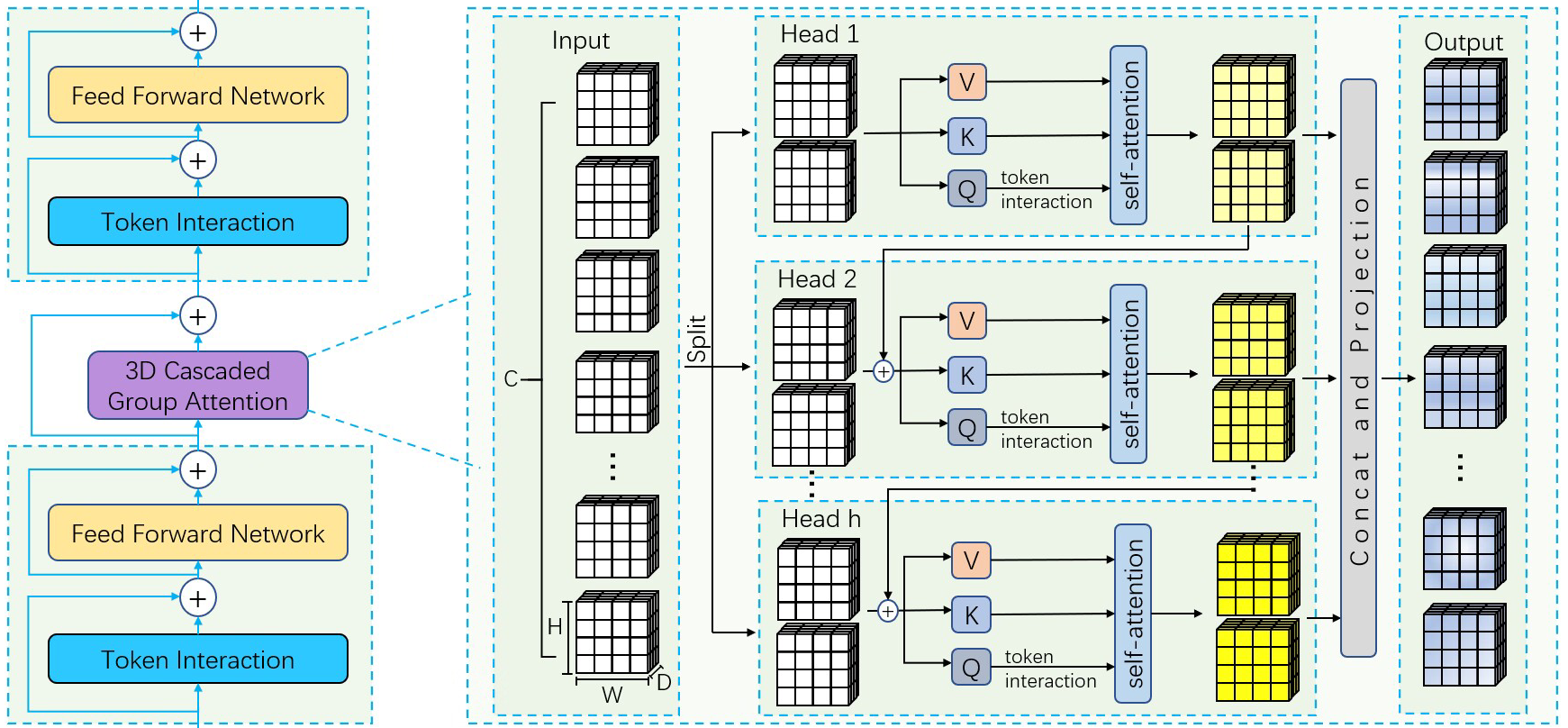}
  \caption{Detailed Structure Display of 3D EfficientViT Block}
  \label{CGA}
\end{figure}

\textbf{Cascaded Group Attention-based 3D EfficientViT Block}: As shown in Fig.\ref{CGA}, the entire 3D EfficientViT \cite{liu2023efficientvit} block presents a "sandwich" layout, using more memory-efficient FFN layers for channel communication and fewer memory-limited self-attention layers. This block contains $N$ FFN layers, before and after every single self-attention layer. The self-attention layer's impact on memory and time consumption is mitigated by this design. Furthermore, an additional token interaction layer using depthwise convolutions (DWConv) \cite{2017MobileNets} is applied before each FFN, thus introducing an inductive bias prioritizes local structural information to enhance the capabilities of the model \cite{2021CoAtNet}. The entire calculation process can be formulated as:
\begin{equation}
\Phi_{i}^{C}(X)=\Phi_{i}^{F}(\Phi_{i}^{T}(X))\label{eq}
\end{equation}
\begin{equation}
X_{i+1}=\prod^{N}\Phi_{i}^{C}(\Phi_{i}^{A}(\prod^{N}\Phi_{i}^{C}(X_{i})))\label{eq}
\end{equation}
where $X_{i}$ is the input of the $i$-th block, and $X_{i+1}$ is the output of the $i$-th block. A $\Phi_{i}^{C}$ block comprises a token interaction layer followed by a FFN layer, symbolized as $\Phi_{i}^{T}$ and $\Phi_{i}^{F}$ respectively. And a single self-attention layer $\Phi_{i}^{A}$ is sandwiched between two $\Phi_{i}^{C}$ blocks. Note that there are residual connections \cite{2016Deep} between adjacent layers.

The attention mechanism the $\Phi_{i}^{A}$ layer is based on is called 3D cascaded group attention (3D CGA), which expressly divides the attention computation into distinct heads by feeding each head with a different split of the entire features. Formally, 3D CGA can be expressed as:
\begin{equation}
\tilde{X}_{ij}=Attn(X_{ij}W_{ij}^{Q},X_{ij}W_{ij}^{K},X_{ij}W_{ij}^{V})\label{eq}
\end{equation}
\begin{equation}
Attn(Q,K,V)=SoftMax(QK^{T}/\sqrt{d}+B)V\label{eq}
\end{equation}
\begin{equation}
\tilde{X}_{i+1}=Concat[\tilde{X}_{ij}]_{j=1:h}W_{i}^{P}\label{eq}
\end{equation}
\begin{equation}
X_{ij}^{'}=X_{ij}+\tilde{X}_{i(j-1)},\quad2\leq j \leq h\label{eq}
\end{equation}
where the 3D input $X_{ij}$ of the $j$-th head (i.e., the $j$-th split of the input feature $X_{i}$) is divided into different subspaces through the projection layers $W_{ij}^{Q}$, $W_{ij}^{K}$, and $W_{ij}^{V}$, and the output $\tilde{X}_{ij}$ is obtained by calculating the self-attention. The self-attention mechanism refers to Swin Transformer \cite{2021Swin1} which is a slight modified of the origin transformer \cite{2017Attention}, where $d$ is the scaling coefficient, and B is trainable bias used as a relative position encoding. The output features of all heads are back to the input dimension through the linear projection layer $W_{i}^{P}$ after concatenating ($h$ is the number of heads). As a cascaded attention mechanism, except for the last head, each head's output is added to the heads that come after it as new input features $X_{ij}^{'}$ instead of $X_{ij}$ to calculate self-attention. 

\noindent\textbf{\large Network Architecture}

\begin{figure}[t]
  \centering
  \includegraphics[width=\textwidth]{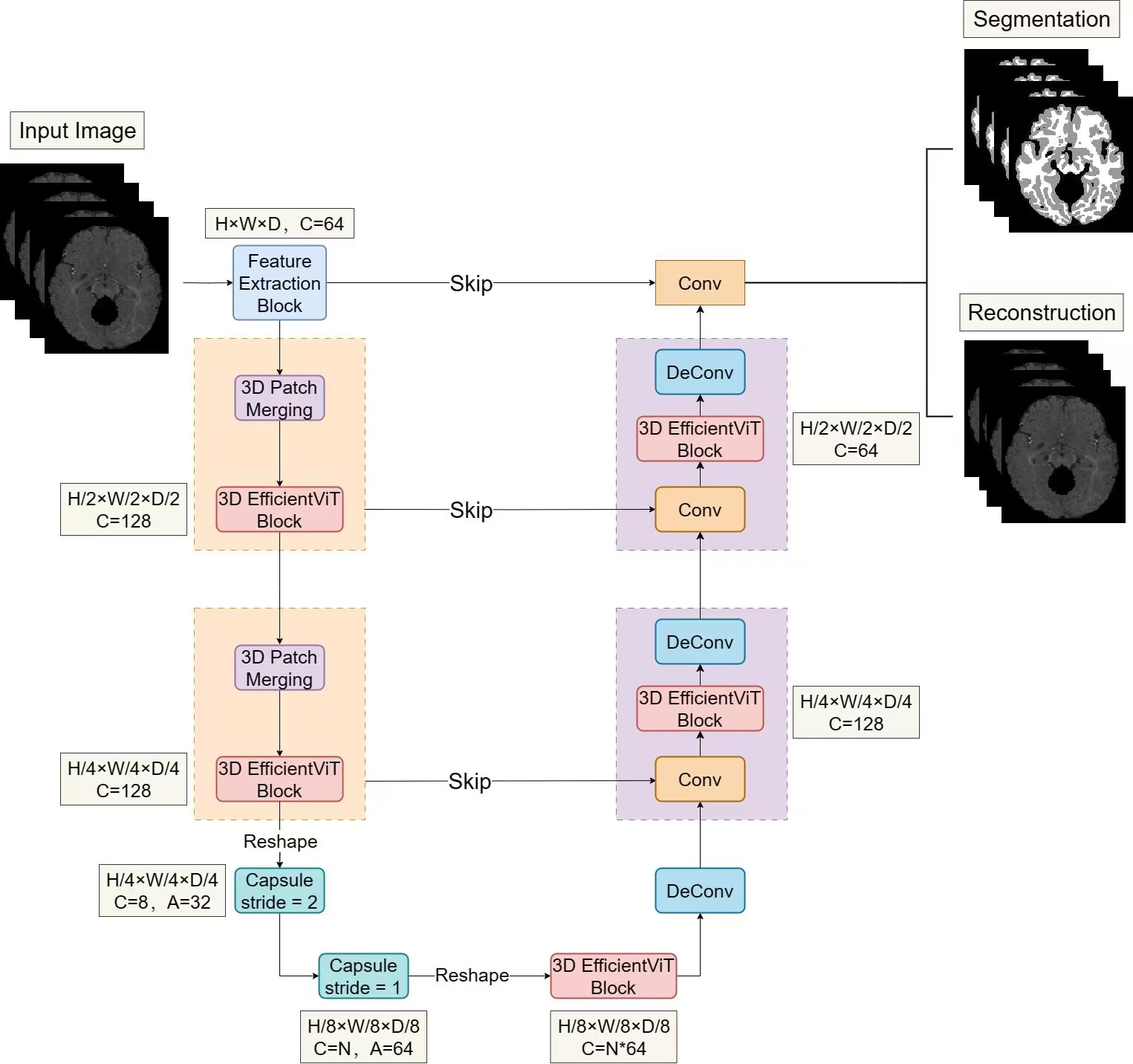}
  \caption{Overview of our proposed 3D-EffiViTCaps, number beside the blocks indicates size of the feature map.}
  \label{Network Architecture}
\end{figure}

The overall network architecture of our proposed 3D-EffiViTCaps is shown in Fig.\ref{Network Architecture}. As a U-Net-shaped \cite{2015U} network, 3D-EffiViTCaps consists of visual feature extraction block, encoder, bottleneck, decoder and skip connections. The input 3D images first pass through the feature extraction block to increase the channel dimension to be processed by the subsequent 3D EfficientViT blocks and capsule blocks. For the encoder, the 3D Patch Merging blocks performs a downsampling operation on the extracted visual features, reducing the resolution of the features and continuing to increase the channel dimension. The 3D EfficientViT blocks joinly capture local and global information and enhances visual feature representation, and are applied to the encoder, bottleneck as well as decoder. The capsule blocks for the bottom layers of the encoder and the bottleneck are used to better learn the long-range dependencies of part-whole relationship. For the decoder, Apart from 3D EfficientViT blocks, we still adopt 3D convolutional blocks like \cite{20223D}, \cite{2022SS}, \cite{20223DConvCaps} as it has been proven in 3D-UCaps \cite{20223D} that using capsules in the upsampling stage will bring huge computational complexity. We also do not use the 3D Patch Expanding blocks as \cite{2021Swin2} and \cite{peiris2022robust} do, as we find that using them in our model would cause performance degradation to a great extent. The specific composition and function of (1) visual feature extraction block, (2) encoder, (3) bottleneck and (4) decoder are as follows:

\textbf{Visual Feature Extraction Block}: In this block, the input volume images are processed by three successive atrous convolutional layers, which transform input images to high-dimentional features so that the ensuing 3D EfficientViT blocks and capsule blocks can process them. Specifically, we use three atrous convolutional layers, setting their output channel numbers to 16, 32 and 64, dilate rates 1, 3, and 3 and kernel size all $5\times 5\times 5$ respectively, with proper padding ensuring that their $H$, $W$, and $D$ dimensions remain unchanged. Thus, the size of the original input image is $H\times W\times D\times C$, and that of the output feature map is $H\times W\times D\times 64$.

\textbf{Encoder}: The encoder receives the visual feature of $H\times W\times D\times 64$ previously transmitted from the feature extraction block. For the low-level feature part of the encoder, we use two 3D Patch Merging blocks to reduce the resolution of the features while basically not losing any information. Each Patch Merging block is followed by a 3D EfficientViT block to gather information on a local and global scale and improve the visual representation of features. Passing through the 3D EfficientViT block, the resolution and channel dimension of the visual features remain unchanged. For the high-level feature part of the encoder, the features are first grouped by channels into a grid of $H\times W\times D$ capsules, and then the 3D capsule layer with a stride of 2 is used for the last downsampling and capture of long-range information. Finally, the semantic information with feature map size $H/8\times W/8\times D/8\times C\times A$ is sent to bottleneck, where $C$ is the number of capsule types and $A$ is the number of each type of capsule (i.e. the dimension of each capsule). 

\textbf{Bottleneck}: We adopt a 3D capsule block and a subsequent 3D EfficientViT block to construct the bottleneck to learn deep visual feature representation. In the bottleneck, the resolution of the feature map remains unchanged, but both $C$ and $A$ of the capsule block are changed, and the size is reshaped to $H/8\times W/8\times D/8\times (N*A)$ after the 3D EfficientViT block to be processed by the convolutional layer of the decoder, where $N$ is the number of categories for classification. To better supervise network training, we use the output of bottleneck to calculate margin loss \cite{2017Dynamic}.

\textbf{Decoder}: Similar to 3D-UCaps \cite{20223D}, we use convolution, deconvolution and skip connections in the decoder pathway. Deconvolution is utilized for upsampling to avoid additional computational cost caused by capsule and the performance loss caused by 3D Patch Expanding. To make up for the information loss that downsampling causes, skip connections are still adopted to fuse the downsampled multi-scale feature passed from the encoder and the upsampled feature obtained through the decoder. In addition to the above blocks, we also add two 3D EfficientViT blocks to enhance the semantic information representation obtained by fusion without significantly increasing the amount of calculation. The resolution of the feature is increased through deconvolution, and at last the channel dimension is changed through a convolutional layer. Thus, the size of the feature map is altered to $H\times W\times D\times N$ and output as the segmentation result.

\noindent\textbf{\large Loss Function}

We use three types of losses to supervise the training process of the network, including:

\textbf{Margin loss}: Initially adopted in \cite{2017Dynamic}, the loss $\mathcal{L}_{margin}$ is mainly used at the last layer of downsampling path, and in our model the loss is calculated between the output of the bottleneck and the downsampled ground truth segmentation. Assuming $\hat{y}$ is the predicted label while y is the ground truth label, its calculation formula can be expressed as:
\begin{equation}
\setlength{\abovedisplayskip}{3pt} 
\mathcal{L}_{margin}=\hat{y}\times (max(0,0.9-y))^{2}+0.5\times (1-\hat{y})\times (max(0,y-0.1))^{2}\label{eq}
\end{equation}

\textbf{Weighted cross entropy loss}: The loss $\mathcal{L}_{WCE}$ is appied at the very end of the upsampling path to optimize the entire network.

\textbf{Reconstruction loss}: To regularize the network, we append an extra branch to the decoder at the end to reconstruct the original image to calculate the difference as in \cite{2018Capsules}. We use masked mean-squared error for the reconstruction.

The weighted sum of the three losses which represents the overall loss is shown below:
\begin{equation}
\mathcal{L}_{total}=\mathcal{L}_{margin}+\mathcal{L}_{WCE}+\mathcal{L}_{rec}\label{eq}
\end{equation}

\section{Experimental Results}

\noindent\textbf{\large Datasets}

\textbf{iSeg-2017} \cite{2019Benchmark}: It contains 10 training samples with ground truth and 13 testing samples without ground truth. These data contain high-quality brain structure segmentation covering the entire brain, and each sample includes T1- and T2-weighted MR images. Three categories are as follows: white matter (WM), gray matter (GM) and cerebrospinal fluid (CSF).

\textbf{Hippocampus} \cite{2019A}: It contains 260 training and 130 testing mono-modal MR images. Categories are divided into anterior and posterior based on anatomical location.

\textbf{Cardiac} \cite{2019A}: Similar to Hippocampus dataset \cite{2019A}, it contains 20 training and 10 testing mono-modal MR images, while there is only one category. 

\noindent\textbf{\large Evaluation Setup}

We conduct experiments on various datasets, including iSeg-2017 \cite{2019Benchmark}, Hippocampus \cite{2019A} and Cardiac \cite{2019A}. For iSeg-2017, we followed the experiment setup of \cite{2019Skip}, i.e, using 9 subjects for training and 1 subject (subject \#9) for testing. In addition, on Hippocampus and Cardiac, we adopt 4-fold cross-validation on their training set.

\noindent\textbf{\large Implementation Details}

Based on the experimental setup of 3D-UCaps \cite{20223D}, we use MONAI \cite{MONAI} for data import and preprocessing, and Pytorch-lightning as the pipeline for deep learning. The input volumetric images are normalized to [0, 1]. For iSeg-2017 and Hippocampus, we set the patch size as $32\times 32\times 32$, while for Cardiac we set the patch size as $64\times 64\times 64$. We set the hyperparameters of the network according to \cite{2018Capsules}. We use Adam as the optimizer with an initial learning rate of 0.0001 and a weight decay value of 0.000002 for L2 regularization. The learning rate is decayed by a rate of 0.1 if, after 5,000 iterations, the Dice score on the validation set does not increase. Early stopping is set with a patience of 25,000 iterations. We conduct experiments on GeForce RTX 3050 with 16GB memory.
 
\noindent\textbf{\large Performance and Comparison}

We compare our proposed 3D-EffiViTCaps with other SOTA 3D CNN-based, 3D Capsule-based and 3D Transformer-based models on three datasets. 

For iSeg-2017 \cite{2019Benchmark}, we report Dice score (DSC) on WM, GM and CSF, and find that our proposed 3D-EffiViTCaps outperforms other models. We provide a quantitative comparison of metric between our model and SOTA MIS models in Table \ref{iSeg-2017 Table}. As shown is the table, compared to classic and recent CNN-based models and Transformer-based models, our model performs better overall and on average with a considerable gap. Especially for WM segmentation, the DSC of our model gains nearly 2 percentage points compared with the second-best \cite{khaled2023independent}. Furthermore, when compared to the SOTA model based on 3D capsule \cite{20223DConvCaps}, our 3D-EffiViTCaps perform obviously better. The average DSC has risen by 0.71\%, and the segmentation performance of WM and GM has enhanced even more significantly. 

\vspace{-20pt} 

\begin{table}[h]
\caption{Comparison on iSeg-2017 with 9 subjects for training and subject \#9 for testing. The best is highlighted in bold, while the second best is underlined.}
\begin{center}
\begin{tabular}{cc|c|cccc}
\hline
\multicolumn{2}{c|}{\multirow{2}{*}{Method}}                  & \multirow{2}{*}{Year} & \multicolumn{4}{c}{DSC↑}                                          \\
\multicolumn{2}{c|}{}                                         &                       & WM             & GM             & CSF            & Average        \\ \hline
\multirow{5}{*}{3D CNN-based MIS}      & 3D U-Net \cite{iek20163D}            & 2016                  & 89.83          & 90.55          & 94.39          & 91.59          \\
                                       & 3D-SkipDenseSeg \cite{2019Skip}    & 2019                  & 91.02          & 91.64          & 94.88          & 92.51          \\
                                       & Non-local U-Net \cite{2020Non}     & 2020                  & 91.03          & 92.45          & 95.30          & 92.77          \\
                                       & APRNet \cite{2021APRNet}              & 2021                  & 91.25          & 92.46          & 95.43          & 93.05          \\
                                       & Bhairnallykar et al. \cite{2023DN} & 2023                  & 91.18          & 93.08          & 95.60          & 93.29          \\ \hline
\multirow{2}{*}{3D Capsule-based MIS}  & 3D-UCaps \cite{20223D}            & 2021                  & 90.95          & 91.34          & 94.21          & 92.17          \\
                                       & 3DConvCaps \cite{20223DConvCaps}          & 2022                  & \underline{92.31}    & 92.73          & \textbf{95.62} & \underline{93.56}    \\ \hline
\multirow{3}{*}{3D Transformer-based MIS} & nnFormer \cite{2021nnFormer}           & 2021                  & 91.66          & 91.80          & 94.21          & 92.56          \\
                                       & VT-UNet \cite{peiris2022robust}             & 2022                  & 91.23          & 92.26          & 94.78          & 92.76          \\
                                       & Afifa et al. \cite{khaled2023independent}        & 2023                  & 91.86          & \underline{93.37}    & 95.23          & 93.49          \\ \hline
\multicolumn{2}{c|}{\textbf{Our 3D-EffiViTCaps}}              & -                     & \textbf{93.59} & \textbf{93.61} & \underline{95.61}    & \textbf{94.27} \\ \hline
\end{tabular}
\label{iSeg-2017 Table}
\end{center}
\end{table}

\vspace{-30pt} 

\begin{table}[h]
\caption{Comparison on Hippocampus and Cardiac both with 4-fold.}
\begin{center}
\resizebox{\columnwidth}{!}{%
\begin{tabular}{cccccc|cc}
\hline
\multicolumn{2}{c|}{\multirow{3}{*}{Method}}                             & \multicolumn{4}{c|}{Hippocampus}                                                                               & \multicolumn{2}{c}{Cardiac}                                     \\ \cline{3-8} 
\multicolumn{2}{c|}{}                                                    & \multicolumn{3}{c|}{DSC↑}                                                          & \multirow{2}{*}{Params(M)} & \multicolumn{1}{c|}{DSC↑}            & \multirow{2}{*}{Param(M)} \\
\multicolumn{2}{c|}{}                                                    & Anterior       & Posterior                  & \multicolumn{1}{c|}{Average}        &                            & \multicolumn{1}{c|}{Left Atrium}    &                           \\ \hline
\multirow{4}{*}{3D CNN-based MIS}      & \multicolumn{1}{c|}{DeepLabv3+ \cite{2018Encoder}} & 87.52          & 85.88                      & \multicolumn{1}{c|}{86.70}          & 5.22                       & \multicolumn{1}{c|}{90.00}          & 5.22                      \\
                                       & \multicolumn{1}{c|}{UNet++ \cite{2018UNet}}     & \underline{88.78}    & \underline{86.98}                & \multicolumn{1}{c|}{\underline{87.88}}    & 2.21                       & \multicolumn{1}{c|}{91.38}          & 49.35                     \\
                                       & \multicolumn{1}{c|}{Unet 3+ \cite{2020UNet}}    & 86.27          & 83.72                      & \multicolumn{1}{c|}{85.00}          & 2.07                       & \multicolumn{1}{c|}{65.55}          & 18.13                     \\
                                       & \multicolumn{1}{c|}{nnU-Net \cite{2021nnU}}    & 88.66          & 86.91                      & \multicolumn{1}{c|}{87.79}          & 1.93                       & \multicolumn{1}{c|}{\underline{91.91}}    & 29.97                     \\ \hline
\multirow{2}{*}{3D Capsule-based MIS}  & \multicolumn{1}{c|}{3D-UCaps \cite{20223D}}   & 85.07          & 82.49                      & \multicolumn{1}{c|}{83.78}          & 3.16                       & \multicolumn{1}{c|}{90.82}          & 2.98                      \\   &
\multicolumn{1}{c|}{3DConvCaps \cite{20223DConvCaps}}        & 88.06                           & 86.47          & \multicolumn{1}{c|}{87.27} &    3.59           & \multicolumn{1}{c|}{90.94}                      & 3.40                                        \\ \hline
\multirow{2}{*}{3D Transformer-based MIS} & \multicolumn{1}{c|}{nnFormer \cite{2021nnFormer}}  & 88.75          & 86.63                      & \multicolumn{1}{c|}{87.69}          & 9.34                       & \multicolumn{1}{c|}{91.75}          & 9.34                      \\
                                       & \multicolumn{1}{c|}{VT-UNet \cite{peiris2022robust}}    & 87.36          & 86.58                      & \multicolumn{1}{c|}{86.97}          & 5.14                       & \multicolumn{1}{c|}{91.88}          & 5.14                      \\ \hline
\multicolumn{2}{c|}{\textbf{Our 3D-EffiViTCaps}}                         & \textbf{89.03} & \textbf{87.20}             & \multicolumn{1}{c|}{\textbf{88.12}} & 4.07                       & \multicolumn{1}{c|}{\textbf{92.74}} & 3.69                      \\ \hline
\end{tabular}
}
\label{Hippocampus and Cardiac Table}
\end{center}
\end{table}

\vspace{-20pt} 

As for results on Hippocampus \cite{2019A} and Cardiac \cite{2019A}, we show them in Table \ref{Hippocampus and Cardiac Table}. We can observe that for both datasets, 3D-EffiViTCaps has a greater advantage than the 3D Transformer-based SOTA 3D MIS model nnFormer \cite{2021nnFormer} and VT-UNet \cite{peiris2022robust} in terms of performance and parameter efficiency. For the powerful CNN-based models UNet++ \cite{2018UNet} and nnU-Net \cite{2021nnU}, 3D-EffiViTCaps leads the former by 0.24\% and 1.36\%, and the latter by 0.33\% and 0.83\% in average DSC of the two datasets, respectively. Besides, the parameter quantities do not vary hugely from dataset to dataset as they do. Finally, on these 2 datasets, 3D-EffiViTCaps significantly surpassed the SOTA model 3D-ConvCaps \cite{20223DConvCaps} by 0.85\% and 1.80\% accordingly with an acceptable increase in parameters.

\noindent\textbf{\large Ablation Study}

We further conduct ablation experiments on the blocks we adopt in the encoder, bottleneck, and decoder of our proposed 3D-EffiViTCaps. Experiments are conducted on Hippocampus \cite{2019A} and we report the DSC, Precision and Recall metrics, as well as the number of parameters and FLOPs of the model. We take the following measures to verify the rationality of the settings of each block with 3DConvCaps \cite{20223DConvCaps} as the baseline (comparing changes in model performance and efficiency): (1) using convolution with stride 2 instead of Patch Merging, (2) using Patch Expanding blocks instead of deconvolution, (3) without 3D EfficientViT (abbreviated as EV in the table) block in bottleneck, (4) without 3D EfficientViT block in decoder and (5) increasing channel dimension to 256 (default 128) after second Patch Merging block. The results are presented in Table \ref{Ablation Study Table}.

\vspace{-20pt} 

\begin{table}[h]
\caption{Ablation Study for 3D-EffiViTCaps on Hippocampus}
\begin{center}
\begin{tabular}{c|ccc|cc}
\hline
\multirow{2}{*}{Method}       & \multicolumn{3}{c|}{Anterior/Posterior}                                                                                                                                                         & \multirow{2}{*}{\begin{tabular}[c]{@{}c@{}}Params\\ (M)\end{tabular}} & \multirow{2}{*}{\begin{tabular}[c]{@{}c@{}}FLOPs\\ (G)\end{tabular}} \\
                              & Recall↑                                                       & Precision↑                                                    & DSC↑                                                            &                                                                       &                                                                      \\ \hline
Baseline                      & \multicolumn{1}{c|}{88.66/85.36}                              & \multicolumn{1}{c|}{87.9/\textbf{87.94}}     & 88.06/86.47                                                     & 3.59                                                                  & 32.30                                                                \\
\textbf{Our 3D-EffiViTCaps}   & \multicolumn{1}{c|}{\underline{89.39}/87.60} & \multicolumn{1}{c|}{89.01/\underline{87.17}} & \textbf{89.03}/\textbf{87.20} & 4.07                                                                  & 33.86                                                                \\
Using convolution stride 2    & \multicolumn{1}{c|}{\textbf{90.88}/86.93} & \multicolumn{1}{c|}{87.35/87.14}                              & \underline{88.93}/86.85                        & 4.53                                                                  & 35.44                                                                \\
Using Patch Expanding         & \multicolumn{1}{c|}{86.71/\textbf{88.03}}    & \multicolumn{1}{c|}{\textbf{90.09}/86.33}    & 88.04/86.95                                                     & 3.39                                                                  & 33.15                                                                \\
w/o 3D EV block in bottleneck & \multicolumn{1}{c|}{89.38/\underline{87.68}} & \multicolumn{1}{c|}{88.58/86.79}                              & 88.78/\underline{86.96}                        & 3.70                                                                  & 33.82                                                                \\
w/o 3D EV block in decoder    & \multicolumn{1}{c|}{88.15/87.34}                              & \multicolumn{1}{c|}{\underline{89.92}/86.74} & 88.85/86.81                                                     & 3.84                                                                  & 32.91                                                                \\
128 channels → 256 channels   & \multicolumn{1}{c|}{89.16/86.87}                              & \multicolumn{1}{c|}{88.93/86.83}                              & 88.84/86.62                                                     & 8.08                                                                  & 45.96                                                                \\ \hline
\end{tabular}
\label{Ablation Study Table}
\end{center}
\end{table}

\vspace{-20pt} 

We can see that compared with the baseline, the performance of the models in each set of experiments using different methods has been improved to a certain extent, owing to our improvements to the model. Among them, we can see that our 3D Patch Merging block can improve model efficiency and performance at the same time. Although using the Patch Expanding blocks can reduce the number of model parameters, it will greatly reduce the segmentation performance of the model. The results of methods (3) and (4) show that adding the 3D EfficientViT block to the bottleneck and decoder can improve certain model performance while only increasing the calculation amount slightly. Furthermore, the model using method (5) will not only reduce the performance to varying degrees, but also increase the amount of parameters and FLOPs. Therefore, we use deconvolution instead of the Patch Expanding block in the decoder, and do not increase the channel dimension in the second Patch Merging block. In addition, we also add 3D EfficientViT blocks in both the bottleneck and decoder.

\begin{figure}[t]
  \centering
  \includegraphics[width=\textwidth]{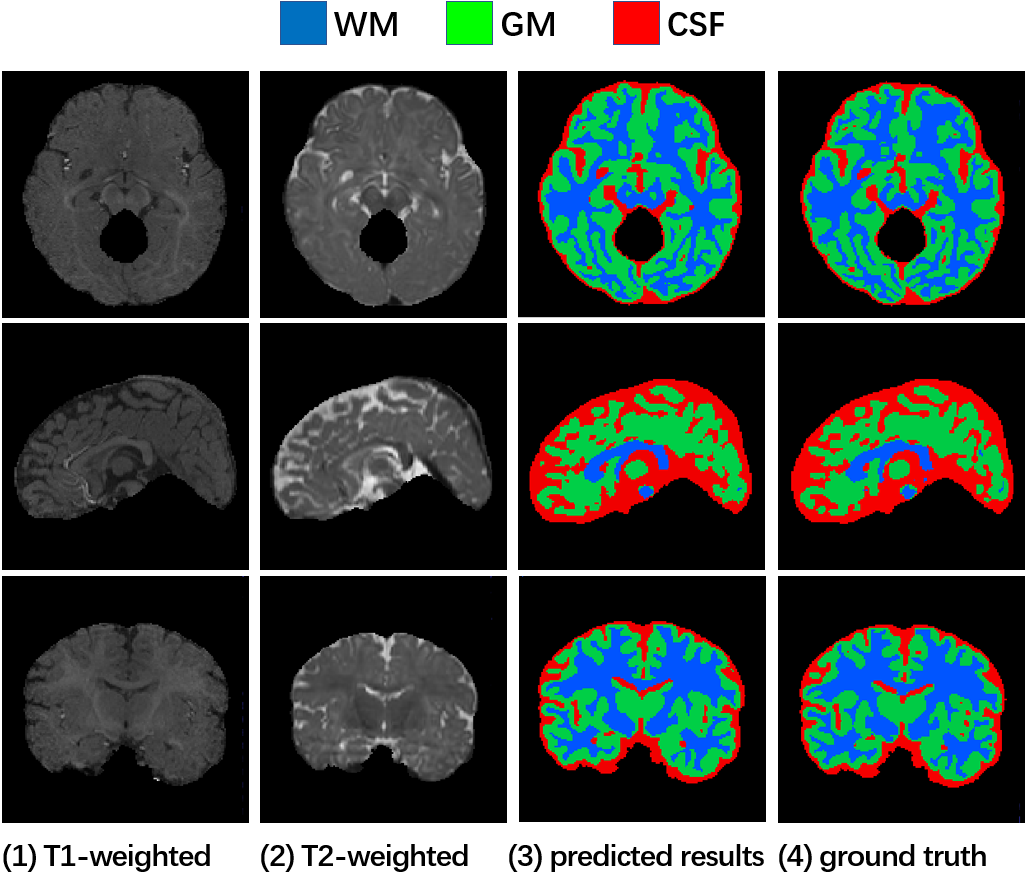}
  \caption{Visualization of segmentation results of subject \#9 in iSeg-2017. Top-down: visualization in three separate planes: axial, sagittal, and coronal.}
  \label{Visualization}
\end{figure}

In addition, we take subject \#9 in iSeg-2017 \cite{2019Benchmark} as an example to visualize the segmentation results of our proposed 3D-EffiViTCaps, which is shown in Fig.\ref{Visualization}. The figure shows that the segmentation results of our 3D-EffiViTCaps are quite similar to the ground truth.

\section{Conclusion}
In this paper, we proposed a U-shaped 3D MIS model that incorporates the merits of CNN, capsule, and EfficientViT. We make full use of the 3D EfficientViT blocks to joinly capture local and global information and enhance visual feature representation, while utilizing the capsule blocks to better learn the long-term dependencies of part-whole relationship. Our experiments show that it outperforms previous SOTA 3D CNN-based, 3D Capsule-based, and 3D Transformer-based models in terms of robust segmentation performance, along with a good balance between performance and efficiency. However, our model falls short of other efficient network models with respect to efficiency. Thus, for future work, we will consider making the network more lightweight by, for example, rethinking the network structure while attempting to maintain model performance. In addition, since there is a large amount of unlabeled data in the medical image segmentation datasets, we will explore self-supervised learning and other methods in an effort to enhance the model's capacity for feature extraction.

\bibliographystyle{splncs04}
\bibliography{refs}

\end{document}